\documentclass[prb,twocolumn,showpacs,amsmath,amssymb,aps]{revtex4}

\usepackage{graphicx}
\usepackage{dcolumn}
\usepackage{amsmath}

\usepackage[format=plain,justification=centerlast]{caption}
\begin{document}

\title{Magnetic field dependent impact ionization in InSb }

\author{Jinki Hong$^{1}$}   
\author{Taeyueb  Kim$^{1,2}$}
\author{Sungjung  Joo$^{1,2}$}
\author{Jin Dong Song$^3$  }
\author{Suk Hee Han$^2$}
\author{Kyung-Ho Shin$^2$  }
\author{Joonyeon Chang$^2$  } \email{  presto@kist.re.kr  }

\affiliation{ $^{1}$Department of Physics, Korea University,
 Chochiwon  339-700, Korea\\
$^{2}$Spin Convergence Research Center, KIST, Seoul 130-650, Korea
\\
$^{3}$Nano Photonics Research Center, KIST, Seoul 136-791 KOREA
 }


\begin{abstract}
Carrier generation by impact ionization and subsequent recombination
under the influence of magnetic field has been  studied for InSb
slab. A simple analytic expression for threshold electric field  as
a function of magnetic field     is     proposed.
Impact ionization is suppressed by  magnetic field.
However, surface recombination is dependent on the polarity of
magnetic field: strengthened in one direction and suppressed on the
opposite direction.
The former contributes quadratic increase to threshold electric
field, and the latter gives   additional linear dependence on
magnetic field.
Based on this study, electrical switching devices driven by magnetic
field can be designed.

\end{abstract}



\maketitle


\begin{figure*}
\centering
\includegraphics[width=0.7\textwidth]{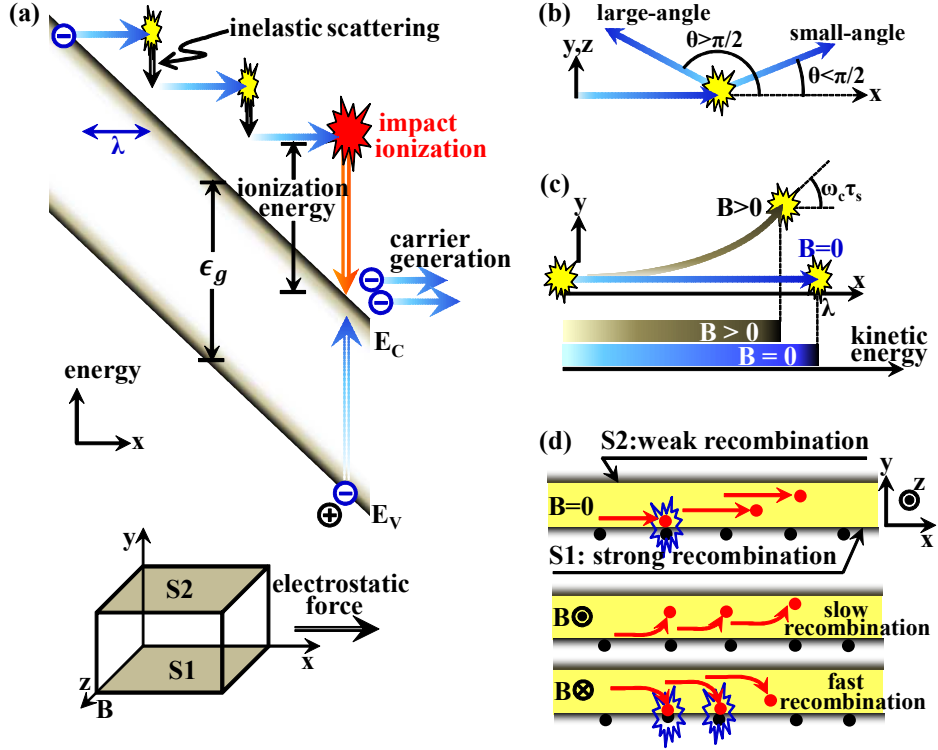}
\captionsetup{ width=1.0\textwidth} \caption{(Color online)
Schematic diagrams illustrating  transport mechanism in impact
ionization process.
(a) An electron accelerated by electric field achieves its kinetic
energy equal to ionization energy and generates electron-hole pair.
Before reaching  ionization energy, the electron experiences
inelastic scattering. (b) Inelastic scattering can be classified to
two groups: small- and large- angle scattering.
(c) Between the scatterings, the trajectory of the electron is
deflected by magnetic field $B$, which suppresses acquisition of
kinetic energy and therefore impact ionization.
(d) Recombination is dependent on the polarity of magnetic field
when recombination strength at S1 and S2 are different.
Axis and two interface, S1 and S2, in the sample are depicted in the
bottom of (a).
 } \label{fig.1}
\end{figure*}


There are many reports about the impact ionization in
semiconductors. However, it is hard to find the one explaining the
influence of magnetic field on the impact ionization. To the best of
our knowledge it is only recently that a reasonable pilot model has
been proposed for the magnetic field effect and compared with
experimental results \cite{APL-MCT}. The previous model is
restricted to a special case: electric transport is quasi-ballistic,
and carrier recombination is independent of magnetic field. We
approach to this issue from more general background, in which an
electron can experience many scatterings before reaching the impact
ionization, and magnetic field affects carrier recombination
process.

Our model gives a result that magnetic field contributes to the
carrier generation and recombination process: the field reduces the
generation rate and increases the threshold voltage, and it also
makes recombination rate sensitive to the polarity of magnetic
field. After describing   some picture regarding our model
qualitatively, quantitative treatment will be followed.



When a high bias voltage is applied, electrons accelerate to a high
speed. If the kinetic energy acquired from the electric field equals
the ionization energy, impact ionization occurs. Upon impact with
the lattice, the electron expends its kinetic energy on ionizing a
valence electron (refer Fig. 1 (a)). This process produces
electron-hole pairs and abruptly increases the electric current in
the device. Impact ionization makes equal number of excess electrons
and holes. Because electron mobility is more than 100 times larger
than that of hole in InSb, we consider only electronic conduction in
impact ionization regime. Before reaching the ionization energy, the
energetic electron can experience energy loss due to inelastic
scatterings. To achieve impact ionization, the electron should
accumulate kinetic energy despite the inelastic scatterings.
Magnetic field affects this carrier generation process. When
magnetic field is applied, the Lorentz force deflects the electronic
trajectory, and the net gain of kinetic energy for a given path
length is reduced (Fig. 1 (c)). To achieve the ionization energy a
longer trajectory is required; however, this longer trajectory gives
rise to the greater possibility of inelastic scattering. Thus, the
deflection of the electronic trajectory caused by magnetic field
leads to suppression of the impact ionization. To restore impact
ionization, a greater electric field  is needed to increase the net
energy gain between the scatterings. Consequently magnetic field
suppresses the carrier generation and increases the threshold
electric field.

Recombination is an elimination process of electron-hole pairs and
generally follows carrier generation. We are interested in
recombination at the two interface, S1 and S2, as depicted in Fig.
1(d). S1 has higher recombination velocity than that of S2 and
carrier electrons are readily recombined near S1. Magnetic field
produces the Lorentz force.  Electrons accumulate or deplete near S1
according to the polarity of magnetic field. When  the polarity of
magnetic field is negative, negative z-direction in Fig. 1(d), the
Lorentz force deflects electrons to S1 and  recombination process is
facilitated, whereas positive polarity makes electrons near S1
depleted and results in  relatively slow recombination.

When the bias voltage exceeds  threshold voltage,  generation rate
is larger than  recombination rate, and then  number of electrons
increases with time, which is known as avalanche state. Under
steady-state, however,  generation process is balanced with
recombination, i.e., generation rate is same as recombination rate.
In this work we consider  the limit of steady-state, which is on the
border of avalanche state.



Now we are in position to treat the model quantitatively. The model
proposed in this work presents a simple  analytic expression for the
threshold electric field as a function of magnetic field.
For small band gap semiconductors such as InSb, the ionization
energy is approximately equal to the band gap energy
$\varepsilon_{g}$ \cite{Narrow}. Energetic electrons undergo
inelastic scatterings before their kinetic energies reach
$\varepsilon_{g}$. The dominant scattering process of InSb at room
temperature is optical phonon scattering \cite{Ophonon1,Ophonon2},
and the optical phonon energy $\hbar \omega_{o}$ is known to be 23
meV \cite{Ophonon3}. Hence, in the present model each scattering
makes an energy loss of $\hbar \omega_{o}$. We will obtain a
probability for an electron to acquire $\varepsilon_{g}$ in spite of
energy losses due to inelastic scatterings. Assuming this
probability is proportional to carrier generation rate, generation
rate will be expressed in terms of electric and magnetic field.
Introducing steady-state condition and recombination parameters, a
simple analytic relation between the threshold electric field and
magnetic field  will be proposed.

\begin{figure}
\centering
\includegraphics[width=0.5\textwidth]{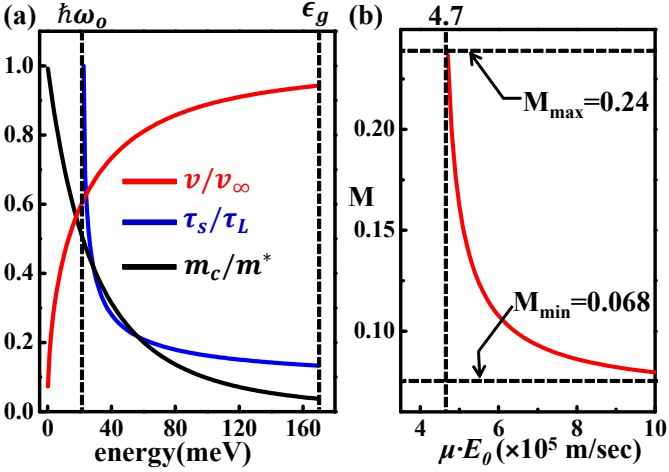}
\captionsetup{ width=0.5\textwidth} \caption{(Color online) (a)  The
parameters in  Eq.~(\ref{eq:Def}) versus energy. The curve of
 $\tau_{s}/\tau_{L}$ indicates that small angle scattering is preferable to
 large angle scattering. $v_{\infty}$ is an asymptotic velocity defined by
$ \sqrt{\varepsilon_{g}/(2 m_{c})}$.
(b) $M$ versus mobility times threshold field $\mu \, E_{0}$.
 The magnitude of $M$ and $\mu \, E_{0}$ are restricted:  $ 0.068 < M <
 0.24$ and $\mu \, E_{0} > 4.7 \times 10^{5} $ m/sec.
 Note that only intrinsic properties of InSb ($\varepsilon_{g}$, $m_{c}$  and $\hbar \omega_{o}$) are used to obtain (a) and (b).
%
%
%
%
%
%
%
} \label{fig.2}
\end{figure}


Adopting Dumke's theory for InSb \cite{Dumke}, scatterings are
classified into two groups according to scattered directions: small-
and large-angle scattering (Fig. 1(b)). For an electron incident
along the direction of an electric force, the small-angle scattering
produces a scattered angle less than $\pi/2$, and the electron is
ready to be accelerated again by the electric field after the
scattering. The large-angle scattering gives a scattered angle
greater than $\pi/2$, which results in deceleration and lose a
chance to acquire further kinetic energy from the electric field.
Thus, large-angle scattering should be avoided to achieve impact
ionization.

Our model starts from this classification. The large-angle
scattering probability in a time interval of $dt$ can be given by
$dt/\tau_{L}$, where $\tau_{L}$ is the relaxation time of the
large-angle scattering. Then, the probability $P$ of an electron
surviving large-angle scattering is $P=exp(-\int 1/\tau_{L}dt)$ for
a finite period of time. $P$ can be expressed with an electric
field. An electric field $E$ supplies energy to an electron at a
rate of $e E \langle v_{x} \rangle$,
where $e$ is the electron's unit charge, and $\langle v_{x} \rangle$
is an average velocity parallel to the electrostatic force in the
time interval between the successive small-angle scatterings, i.e.,
$\langle v_{x} \rangle \equiv ({1 / \tau_{s}}) \int_{0}^{\tau_{s}}
v_{x} dt $, where $\tau_{s}$ is the relaxation time of the
small-angle scattering. Energy loss by the small-angle scattering is
given by $\hbar \omega_{o}/\tau_{s}$. Thus, the net rate of energy
gain  is expressed by $ d\varepsilon/dt= e E \langle v_{x} \rangle -
\hbar \omega_{o}/\tau_{s}$ . Using $dt=d\varepsilon/(e E \langle
v_{x} \rangle - \hbar \omega_{o}/\tau_{s}) $, $P$ is rewritten as
%
%
%
%
      \begin{equation}
      \label{eq:2}
             P=exp\left(
               \int_{\hbar \omega_{o}}^{\varepsilon_{g}}
               -{ 1 \over \tau_{L}} \cdot
                {{d \varepsilon} \over { e E \langle v_{x} \rangle -  \hbar \omega_{o}/\tau_{s}             }}
               \right)
              .
    \end{equation}
A reasonable choice for the lower bound in the integrand is
  $\hbar \omega_{o}$ because optical phonon scattering is absent for an
electronic energy less than $\hbar \omega_{o}$. When an electron
with an initial energy of $\hbar \omega_{o}$ moves to a final energy
states of $\varepsilon_{g}$, various paths are possible, and $P$ is
dependent on these paths. We only consider a path which gives a
maximum value of $P$. The path having the shortest electronic
path-length gives the maximum $P$. A diffusion effect in the energy
space is not considered; it is a second order effect
\cite{EnergyDiffusion} and ignored in this work.

\begin{figure}
\center
\includegraphics[width=0.4\textwidth]{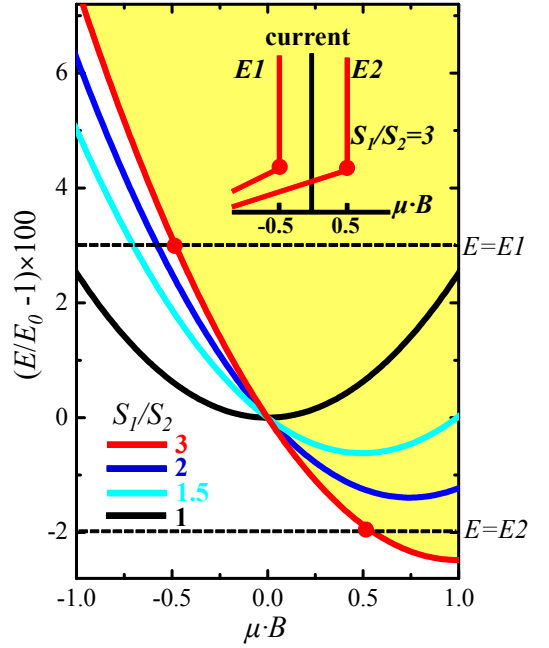}
\captionsetup{ width=0.5\textwidth} \caption{(Color online)
Normalized threshold electric field $E/E_{0} -1$  as a function of
mobility times magnetic field $\mu \, B$ for a variety of ratios of
surface recombination velocity, $s_{1}/s_{2}$.
Eq.~(\ref{eq:CR}) is used on the assumption that $\mu \, E_{0}= 5
\times 10^{5} $ m/sec, $s_{2}=10^{4}$ m/sec and $\mu_{p}/\mu$ is
effective mass ratio of electron and hole, 0.03.
The yellow area represents avalanche state for $s_{1}/s_{2}=3$. The
avalanche state is achieved for an given electric field $E=E1$
($E=E2$)   when $\mu \, B > -0.5$ ($\mu \, B > 0.5$). This
characteristic provides abrupt change of current in current versus
$\mu \, B$ plot.
The inset shows an expected schematic trace for $E=E1$ and $E2$,
which can be used as an electrical switch controlled by magnetic
field.
 } \label{fig.3}
\end{figure}


To elucidate $\langle v_{x} \rangle$ in Eq.~(\ref{eq:2}), more
details for the small-angle scattering are needed. Velocity is
generally dependent on kinetic energy. However, due to the band
non-parabolicity \cite{Kane} of InSb, the velocity of InSb is a slow
function of energy for energetic electrons (refer Fig. 2(a)). Hence,
the variation of the magnitude of velocity between the successive
small-angle scatterings is considered to be negligibly small even
though the corresponding energy change is considerable. After a
scattering event the electron may have various scattered directions,
and these directions work again as incident directions for the next
scattering. The shortest path is achieved when this scattered
direction is in parallel with the electric force (along x-axis).
Thus, the direction of electronic velocity just after the
small-angle scattering is considered to be in parallel with the
electric force.

Magnetic field is involved with $\langle v_{x} \rangle$ in
Eq.~(\ref{eq:2}). In the presence of a magnetic field $B$, classical
trajectory of an electron is governed by the Lorentz force. For an
electron with velocity of $v$ moving toward the
electric-force-direction (refer Fig. 1(c)), the classical trajectory
gives $1/\tau_{s} \int_{0}^{\tau_{s}} v_{x} dt$
 $=v sin(\omega_{c} \tau_{s})/{\omega_{c} \tau_{s}}$, where
 $\omega_{c} \equiv e B / m^{*}$ and $m^{*}$ is electronic effective mass.
Using the approximation $sin(\omega_{c} \tau_{s})\doteq 1- {1/6}
\cdot (\omega_{c} \tau_{s})^{2}$ for low magnetic field,
$\langle v_{x} \rangle$ in Eq.~(\ref{eq:2})   can be replaced with $
v~ ( 1- {1/6} \cdot (\omega_{c} \tau_{s})^{2})$.
The total scattering rate $1/\tau_{s} + 1/\tau_{L}$ is nearly
independent of energy \cite{Turkey}, and $\tau_{s}$ is favored over
$\tau_{L}$ in the overall energy range (Fig. 2(a)). Thus, electronic
mobility $\mu$ is determined by $\tau_{s}$, which allows  $\tau_{s}=
\mu \, m_{c}/e$. $m_{c}$ is the effective mass at the conduction
band edge.
We consider $E$ as a small deviation from the threshold electric
field at zero magnetic field,  $E_{0}$. The right term in
Eq.~(\ref{eq:2}) can be expanded to the first order of $(E/E_{0} -1)
- {1/6}\cdot (\omega_{c} \tau_{s})^{2}$. Following some algebra, it
is readily shown that
%
%
      \begin{equation}
      \label{eq:PP0}
                  P/P_{0}-1 = C_{G}((E/E_{0} -1) - (1/6) M(\mu \, B )^{2} )
           ,
     \end{equation}
where
%
%
      \begin{eqnarray}
      \label{eq:Def}
             C_{G} & \equiv & \int_{\hbar \omega_{o}} ^{\varepsilon_{g}}
             {\tau_{s}\over \tau_{L}} \cdot
             {{m_{c} \, \mu \, E_{0} \, v }\over (m_{c} \, \mu \, E_{0} \, v   -  \hbar \omega_{o}  )^{2}}
              d\varepsilon
             ,
           \\
             M & \equiv & { 1 \over C_{1}}
             \int_{\hbar \omega_{o}} ^{\varepsilon_{g}}
              {\tau_{s}\over \tau_{L}} \cdot
              {{m_{c} \, \mu \, E_{0} \, v}\over (m_{c} \, \mu \, E_{0} \, v   -  \hbar \omega_{o}  )^{2}}
             \cdot \left(   m_{c} \over m^{*} \right)^{2}
          d\varepsilon
          ,
           \nonumber
    \end{eqnarray}
and $P_{0}$ is amount of $P$ at $E=E_{0}$ and $B=0$, i.e.,
$ exp\left( - \int_{\hbar \omega_{o}}^{\varepsilon_{g}} {
                \tau_{s} \over \tau_{L}} \cdot
                {{d\varepsilon} \over { m_{c} \, \mu \, E_{0} \, v -  \hbar \omega_{o}
                } }
                    \right)
 $.

The parameters in Eq.~(\ref{eq:Def}) can  be calculated as follows.
An analytic form of $\tau_{s}/\tau_{L}$ is given by
$ Log(     (1+k_{r} )^{2}   /  (1+k_{r}^{2} )           )
 /Log(     (1+k_{r}^{2})    /  (1-k_{r}     )^2         )
$
\cite{Dumke}, where $k_{r} \equiv k(\varepsilon) - \hbar
\omega_{o}/k(\varepsilon)$ and $k(\varepsilon)$ is a wave vector of
an electron for a given energy $\varepsilon$. The non-parabolic
dispersion is given by
$\varepsilon(k)=-\varepsilon_{g}/2+ \sqrt{\varepsilon_{g}^{2} /4
 + \varepsilon_{g} \hbar^{2}k^{2}/(2 m_{c})}$
from which the following expressions can be obtained:
$ k(\varepsilon)  = (\sqrt{2 m_{c}}/\hbar)
\sqrt{\varepsilon^{2}/\varepsilon_{g}  + \varepsilon   } ,$
$ v=(1/\hbar)\partial\varepsilon/\partial k$
$=(1/\sqrt{2 m_{c}})\sqrt{\varepsilon^{2}\varepsilon_{g} +
\varepsilon\varepsilon_{g}^{2}  }/(\varepsilon + \varepsilon_{g}/2)
$
and
$ m_{c}/m^{*} \equiv (m_{c}/\hbar^{2} ) \partial^{2}\varepsilon /
\partial^{2} k  = (2 \varepsilon/\varepsilon_{g}  + 1)^{-3}$.
In InSb $\varepsilon_{g}$ is 0.17 eV and $m_{c}$ is  0.013 times the
mass of a free electron.
Using these relations, $C_{G}$ and $M$ in Eq.~(\ref{eq:Def}) can be
obtained.

It is only a few electrons which survive large-angle scattering and
obtain kinetic energies equal to $\varepsilon_{g}$ and finally
contribute to the impact ionization \cite{Shockley}. Because the
impact ionization is governed by the energy-gain process specified
by $P$, a generation rate per unit carrier due to the impact
ionization is assumed to be proportional to $P$. Therefore, $P/P_{0}
-1$ in Eq.~(\ref{eq:PP0}) represents a normalized generation rate.


Recombination relies on magnetic field when the recombination
strength at the two interfaces are different (see Fig. 1 (d)).
In low magnetic field regime the first order approximation of
recombination rate $R(B)$ with respect to $B$ gives ${R(B)/R_{0}} -
1= - C_{R} \, \mu \, B $, where $R_{0}$ is recombination rate at
zero field \cite{PCexp}.
When surface recombination is dominant over bulk recombination and
sample thickness is smaller than carrier diffusion length, the
following simple  expression can be obtained from Lile's
results\cite{Lile}:
%
      \begin{equation}
      \label{eq:CR}
             C_{R}  =    \left(  {1 \over s_{2}} -  { 1 \over s_{1}}
             \right) \mu_{p} E_{0}
             .
    \end{equation}
$s_{1}$ and $s_{2}$ are surface recombination velocities at the two
interfaces, S1 and S2, respectively,  and difference of them gives
non-zero value of $C_{R}$.
The steady-state condition asserts that the generation rate is equal
to the recombination rate, which leads to $P/P_{0}=R(B)/R_{0}$ and
therefore $P/P_{0} -1 =  - C_{R} \, \mu \, B$.   Then,  the
threshold field in Eq.~(\ref{eq:PP0}) is expressed by
%
      \begin{equation}
      \label{eq:Main}
             E/E_{0} -1 = { 1 \over 6 } M
             (\mu B)^{2} - { C_{R} \over C_{G}}  \mu B
             .
    \end{equation}
%
%
%


Plotting threshold field $E$ according to magnetic field permits an
overview of the present model.
The plots  in Fig. 3 are calculated
ones using Eq.~(\ref{eq:CR}) and Eq.~(\ref{eq:Main}). SI unit for
mobility and magnetic field makes $\mu \, B$ dimensionless.
For identical surface recombination at the two interface,
$s_{1}=s_{2}$, the normalized threshold field 
increases quadratically with magnetic field. This increase is caused
by  the suppression of impact ionization by magnetic field.
Difference in the surface recombination velocities, $s_{1} > s_{2}$,
adds the linear term  and the threshold field becomes asymmetrically
dependent  on magnetic field.
Note that the positive (negative) polarity of magnetic field
corresponds to the middle (bottom) diagram in Fig. 1 (d). Weak
(strong) recombination results in small (large) threshold electric
field.

The curves in Fig. 3 represent boundaries between normal and
avalanche states in the space of  electric and magnetic field: the
avalanche and normal states correspond  to the upper and lower areas
of the curve, respectively. The yellow area corresponds to avalanche
state for $s_{1}/s_{2}=3$, for instance.
By varying magnetic or electric field, one of the two conducting
state, normal and avalanche state, can be selected.
An interesting application of this phenomenon  is switching device.
For a given electric field, electric current can be changed abruptly
by varying magnetic field, which can be a good candidate of
magnetic-field-driven electrical switching device (refer the inset
of Fig. 3).






\begin{references}

\bibitem{APL-MCT} J. Lee et al., Appl. Phys. Lett. {\bf 97}, 253505(2010).

\bibitem{Narrow} B.Gelmont, K. Kim and M. Shur, Phys. Rev. Lett. {\bf 69}, 1280
(1992).
\bibitem{Ophonon1} H. Kishan, S. K. Agarwal and K. D. Chaudhuri,
Phys. Rev. B {\bf 28}, 2078 (1983).

\bibitem{Ophonon2} Y. J. Jung, M. K. Park, S. I. Tae, K. H. Lee, and H.
J. Lee, J. Appl. Phys. {\bf 69}, 3109 (1991).

\bibitem{Ophonon3} N. L. Rowell, D. J. Lockwood, G. Yu, Y. Z. Gao, X. Y.
Gong, M. Aoyama and T. Yamaguchi, J. Vac. Sci. Technol. A {\bf 22},
935 (2004).

\bibitem{Dumke} W. P. Dumke, Phys. Rev. {\bf 167}, 783 (1968).

\bibitem{EnergyDiffusion} J. P. Leburton and K. Hess, Phys. Rev. B {\bf 26},
5623 (1982).

\bibitem{Kane} E. O. Kane, J. Phys. Chem. Solids {\bf 1}, 249 (1957).

\bibitem{Turkey} O. Ozbas and M. Akarsu, Turk. J. Phys. {\bf 26}, 283 (2002).

\bibitem{Shockley} W. Shockley, Solid-State Electron. {\bf 2}, 35 (1961).

\bibitem{PCexp} Using photoconductivity method for n-type InSb at room temperature, we observed that
  ${R(B)/R_{0}} - 1$ is  linear with respect to  $B$ and
 $C_{R}=0.11$. This experimental result will be submitted elsewhere.

\bibitem{Lile} D. Lile, J. Appl. Phys. {\bf 41}, 3480 (1970).

\end{references}
\end{document}